\documentclass[runningheads]{llncs}
\usepackage[T1]{fontenc}
\usepackage{graphicx}
\usepackage{subcaption}
\usepackage{hyperref}
\usepackage{fancyhdr}
\usepackage[absolute,overlay]{textpos}
\usepackage{xcolor}

\begin{document}
\begin{textblock*}{\textwidth}(1in, 0.6in)
\centering
  \textbf{Accepted at the NextGen Learning Interfaces Workshop, AIED 2026}
\end{textblock*}

\title{Regulating the AI Tutor: Adolescent Help-Seeking and Self-Regulated Learning with GenAI in Mathematics Learning: A work in progress}

\title{Regulating the AI Tutor: Intentions, Help-Seeking, and Self-Regulated Learning in Adolescent GenAI Use}
\titlerunning{Regulating the AI Tutor}
%
%
\author{Rania Abdelghani\inst{1}\orcidID{0000-0002-6361-6609} \and
Peter Kaiser\inst{1,2}\orcidID{0000-0003-1031-1139} \and
Kou Murayama\inst{1}\orcidID{0000-0003-2902-9600 }}
\authorrunning{R. Abdelghani et al.}
%
\institute{Hector Institute of Education Sciences and Psychology, University of Tübingen, Germany\\
\email{\{rania.abdelghani,pe.kaiser,k.murayama\}@uni-tuebingen.de}
\and
Department of Mathematics, Faculty of Sciences, University of Tübingen, Germany\\
}
\maketitle

\begin{abstract}
Generative AI (GenAI) tools are now common learning companions for adolescents, yet how they regulate their use during authentic learning tasks remains poorly understood. Self-regulated learning (SRL) and high-level help-seeking (HS) are commonly proposed as safeguards against passive or shortcut-oriented use, but most empirical studies focus on aggregate learning outcomes rather than these moment-to-moment processes during AI-supported learning.

This work-in-progress examines open-ended conversational data from 98 Grade-9 students across three German \textit{Gymnasium} schools, who used a web-based Mistral-Large tutor to prepare a curriculum-aligned mathematics skill before an exam. Alongside chat logs (1,616 turns; 808 student turns), we collected pre/post domain knowledge, pre-chat learning needs, and self-reported cognitive load. We propose a turn-level codebook combining theory-driven SRL and HS constructs with two LLM-specific inductive codes (agency over the AI; epistemic vigilance), and report preliminary AI-coded results. 

Although students overwhelmingly selected scaffolded support before the chat, their interactions were dominated by instrumental requests with almost no explicit monitoring or evaluation. Post-test performance was significantly lower than pre-test, and higher extraneous cognitive load predicted lower post-test scores after controlling for prior knowledge. We discuss how these patterns can support hybrid human-AI analysis of interaction patterns and inform scaffolds for more agentic and epistemically proactive GenAI use.
\end{abstract}


\section{Related Work and Current Study}
Generative AI (GenAI) systems are increasingly used by adolescents as everyday learning companions~\cite{hashem2025understanding}. Their fluency and low interactional friction create opportunities for personalized support, but the same affordances may undermine learning when students obtain rapid solutions rather than engage in the effortful reasoning required for conceptual understanding~\cite{wang2025decoupling,darvishi2024impact}. Recent work describes this risk as \textit{cognitive surrender}: adoption of AI-generated reasoning with minimal scrutiny, such that the system suppresses rather than supports thinking~\cite{shaw2026thinking}. 

Self-regulated learning (SRL) theory emphasizes that productive learning requires goal-setting, monitoring, strategy selection, evaluation, and adaptation across forethought, performance, and reflection phases~\cite{Zimmerman2008InvestigatingProspects}. Similarly, help-seeking (HS) research distinguishes \textit{instrumental} help-seeking, in which learners request hints or explanations that preserve responsibility for understanding, from \textit{executive} help-seeking, in which learners obtain completed solutions that bypass learning-relevant effort~\cite{nelson1986help,aleven2003help,tawfik2020role}. Together, SRL and high-level HS provide a useful lens for evaluating whether students use AI as a learning partner or as an answer-producing shortcut. We call this set of skills \textit{epistemic proactivity}: identifying learning goals, deciding how to seek them with AI, evaluating responses, and deciding when further regulation is needed.

However, and in the context of Gen-AI supported learning, the low-friction nature of tools like Large Language Models (LLMs) interaction may destabilize these processes~\cite{tankelevitch2024metacognitive}. In human tutoring, under-specified or premature questions typically elicit clarifying scaffolds. General-purpose LLMs, by contrast, deliver an answer regardless~\cite{cheng2025asking}. Using LLMs while lacking epistemic proactivity may therefore reinforce executive help-seeking, cognitive offloading, and uncritical acceptance of AI output. 

Despite their importance, adolescent SRL and HS strategies with GenAI remain under-explored. Most studies rely on self-report or aggregate pre/post measures, and miss how learning actually unfolds~\cite{Yan2025DistinguishingAI}. A process-sensitive account, grounded in SRL and HS theory, is therefore needed to examine how students regulate their learning with AI as it happens.

The present work-in-progress addresses this gap through a classroom study with Grade-9 students using a web-based Mistral Large tutor during a curriculum-aligned mathematics task. We propose a turn-level coding framework characterizing students' SRL and HS behaviors, mathematical activity, agency over the AI, and epistemic vigilance toward its output. By connecting stated learning goals, interaction patterns, cognitive load, and learning outcomes, the study aims to identify if adolescent learners can transform learning-oriented intentions into productive AI-supported learning.

To address this, we aim to answer the following questions: 1) What SRL and help-seeking behaviors do students enact during the interaction with the AI tutor? And 2) How do students' enacted behaviors align with their stated learning intentions, cognitive load, and learning outcomes?

\section{System and Study Design}
\subsection{Participants}
We recruited 106 Grade-9 students from three public German \textit{Gymnasium} schools in Baden-Württemberg. After excluding 8 for incomplete tasks or questionnaires, the final sample comprised 98 students aged 14--15 ($M=14.67$). 

The study was approved by the University of Tübingen ethics committee and the Ministry of Culture, Youth and Sports Baden-Württemberg. Written informed consent was obtained from students and their legal guardians, and the app included real-time monitoring and safeguards against personally identifiable information, offensive content, and off-task queries.

\subsection{Task Description and Procedure}
\begin{figure}[htbp]
\centering
\includegraphics[width=.9\textwidth]{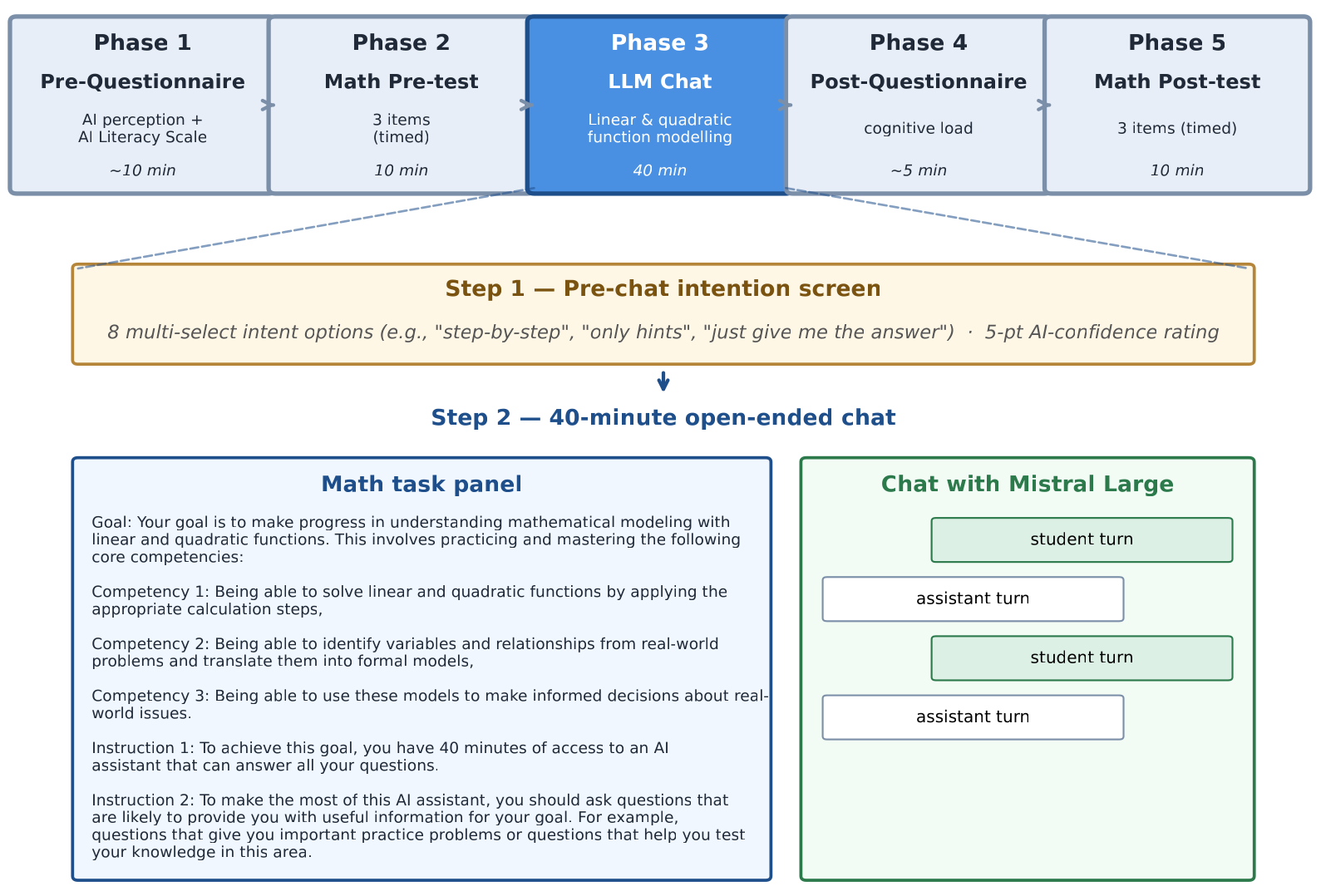}
\caption{Timeline of the study and content of the pedagogical task} 
\label{fig:stud_time_content}
\end{figure}

Students completed a five-phase tablet-based protocol during a regular mathematics lesson (\autoref{fig:stud_time_content}). Mathematics was selected because as it requires several SRL-relevant processes -- goal setting, monitoring, strategic help-seeking, interpretation, and validation~\cite{blum2009mathematical} -- and because it is a domain in which students often experience difficulty, making it an interesting investigation context as it can lead to both productive and shortcut-oriented AI use.

After a short presentation by the research team, students completed a pre-questionnaire including an AI Literacy Scale~\cite{ng2024design}, and a learning-strategies questionnaire~\cite{kim2017establishing}, followed by a timed math pre-test with three validated linear and quadratic items selected for curricular fit. They then interacted with a Mistral Large tutor on a real-world mathematical-modeling task grounded in three worked examples. The task was framed explicitly as an opportunity to practice and make learning progress before an upcoming exam. See~\autoref{fig:stud_time_content}. 

Before starting the chat, students selected the learning goals they hoped to achieve from eight options adapted from question-taxonomy work~\cite{tawfik2020role} -- e.g., concept explanations, strategies, verification of understanding, step-by-step examples, hints without answers, final solutions of examples-- and rated their confidence in achieving them on a 5-point scale. This screen operationalized the forethought phase of the SRL loop. The chat phase ended only after at least six valid task-related turns, with a four-rule gibberish detector excluding empty or non-substantive turns from the count and history. After the chat, students completed six cognitive-load items on a 10-point scale (intrinsic, extraneous, germane)~\cite{Sweller1991EvidenceTheory}, followed by a post-test on mathematical modeling.

The chat phase ended only after students had submitted at least six valid task-related turns. A four-rule gibberish detector excluded empty, irrelevant, and non-substantive turns from the minimum-turn count and from the chat history. After the chat, students completed six cognitive-load items on a 10-point scale, covering intrinsic, extraneous, and germane load~\cite{Sweller1991EvidenceTheory}, followed by a post-test on mathematical modeling. Chat exchanges were stored with timestamps, alongside questionnaire responses, and math answers.


\section{Proposed Method for the Conversational Data Analysis}
\label{sec:codebook-dev}
\begin{figure}
\centering
\includegraphics[width=.7\textwidth]{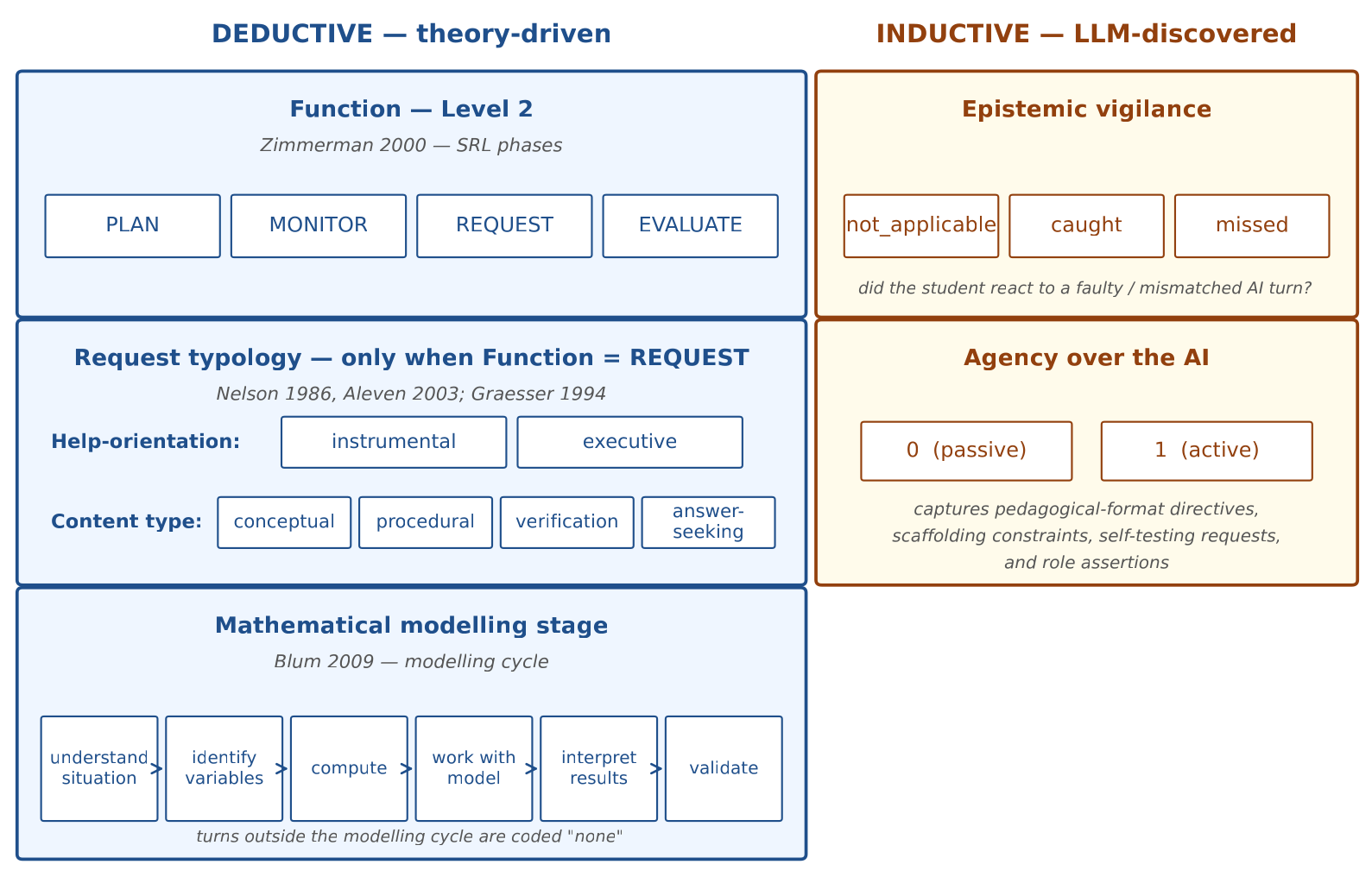}
\caption{Proposed codebook for analyzing students conversational data} 
\label{fig:codebook}
\end{figure}

We propose a hybrid turn-level coding framework combining theory-driven constructs from SRL, HS, and mathematical modeling with LLM-specific inductive constructs (\autoref{fig:codebook}). Coding is applied to each task-relevant student turn.

\paragraph{Deductive codes.} Each turn is coded for its SRL function based on Zimmerman's cyclical model~\cite{Zimmerman2008InvestigatingProspects}: \textit{PLAN} (forethought and task structuring), \textit{MONITOR} (self-diagnosis of comprehension gaps), \textit{REQUEST} (help-seeking), and \textit{EVALUATE} (judgments of the AI's response). REQUEST turns are further coded as \textit{instrumental} or \textit{executive}~\cite{nelson1986help,aleven2003help}, and as \textit{conceptual}, \textit{procedural}, \textit{verification}, or \textit{answer-seeking}~\cite{tawfik2020role}. To capture mathematical quality, we add a modeling-process code based on Blum's framework~\cite{blum2009mathematical}, distinguishing surface-level computation requests from deeper modeling activities (understanding the situation, identifying variables, mathematizing, working with the model, interpreting results, validating).

\paragraph{Inductive LLM-specific codes.}
Two constructs were derived through an exploratory pass with Gemini 2.5 Pro on a randomly selected 30\% of conversations. \textit{Epistemic vigilance} is a three-valued code (\texttt{not\_applicable}, \texttt{caught}, \texttt{missed}) capturing whether a student reacts to an inaccurate or mismatched AI turn. \textit{Agency over the AI} is a binary marker capturing attempts to shape the AI's role, format, or scaffolding (e.g., asking for hints rather than answers, requesting self-testing).

Preliminary results below rely on AI-based coding. A stratified 30\% human-validation pass by mathematics-didactics and learning-sciences experts is ongoing. Future analyses will report per-construct reliability and use the validated scheme to annotate the remaining data.

\vspace{-1cm}

\section{Preliminary Results}
The dataset contains 1,616 chat turns from 98 students, including 808 student turns ($M=8.2$ per session, $Mdn=7$, range $6$--$28$). Because human validation of the codebook is ongoing, behavioral results are provisional AI-coded patterns.

\subsection{Forethought: Students' Stated Learning Intentions}
Before the chat, students selected on average $4.66$ of 8 goals, suggesting broad rather than focused intentions. Choices were strongly learning-oriented: step-by-step examples ($82.9\%$), tips and strategies ($80.3\%$), practice problems ($71.1\%$), concept explanations ($69.7\%$), and checking understanding ($69.7\%$). Only $11.8\%$ selected ``just give me the final solutions'' and $36.8\%$ selected finishing as quickly as possible. Self-reported forethought thus tilted strongly toward scaffolded support rather than answer offloading.

\subsection{Performance: SRL Functions and Help-Seeking}
\begin{figure}%
    \centering
    \subfloat[\centering \textbf{Students mostly used AI to request information} ]{{\includegraphics[width=5cm]{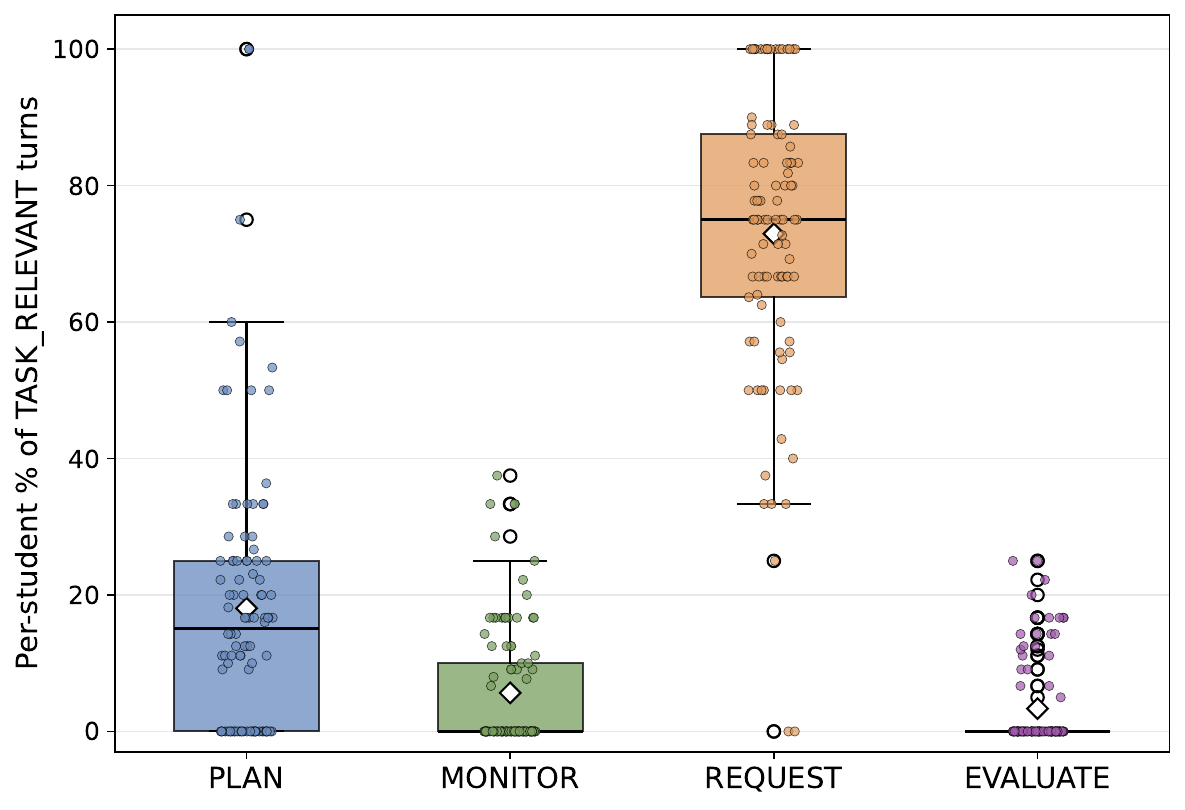} }}%
    \subfloat[\centering \textbf{Students mostly had instrumental HS strategies with the AI}]{{ \includegraphics[width=7cm]{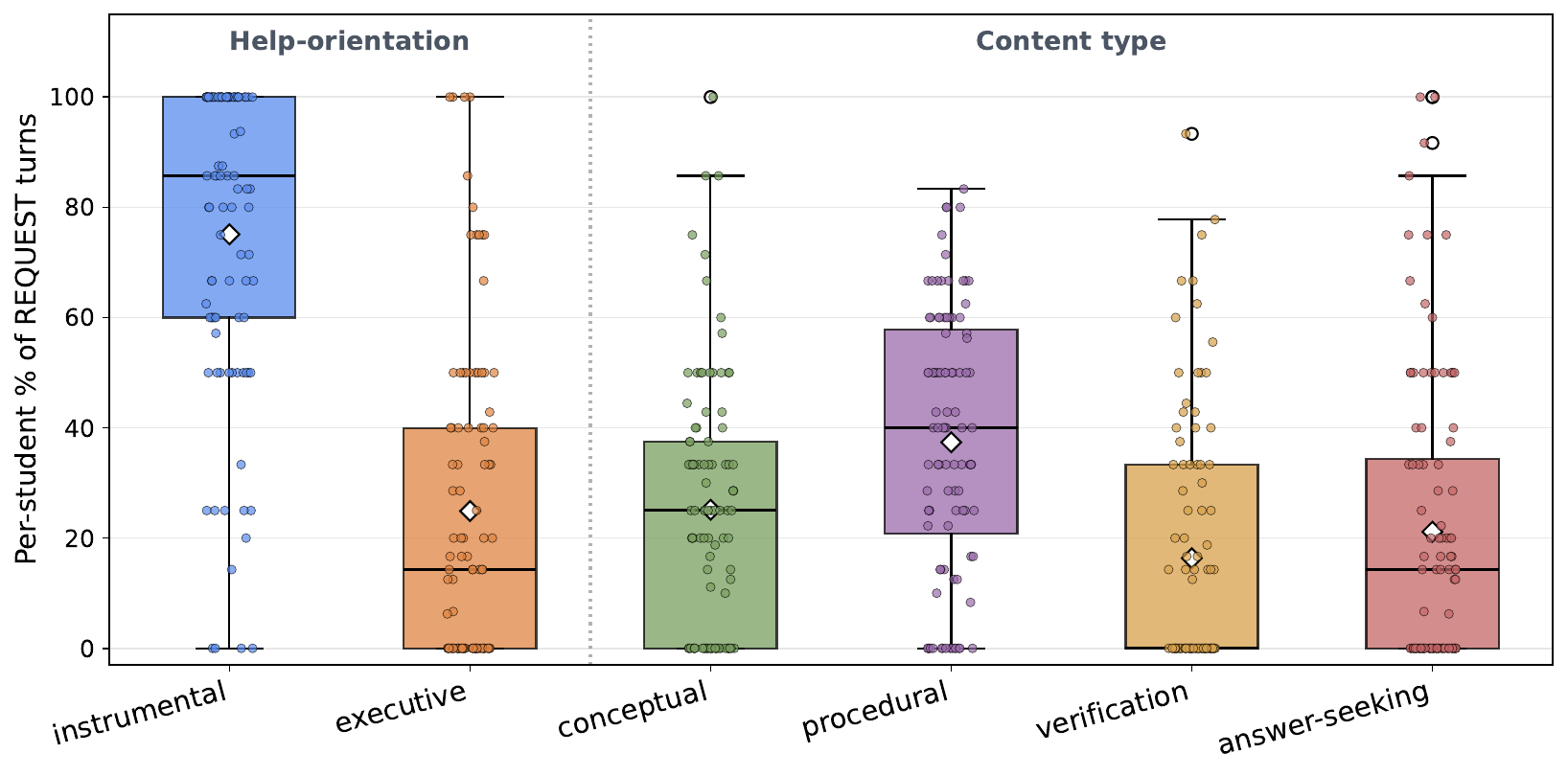} }}%
    \caption{\centering Distribution of students prompts functions and request typology)}%
    \label{fig:res-dist}%
\end{figure}

We applied the deductive scheme to all 808 student turns using Gemini 2.5 Pro (frozen prompt, $T=0$, fixed seed). Interactions were dominated by REQUEST turns ($M=72.9\%$ of task-relevant turns; $Mdn=75.0\%$, IQR $63.6$--$87.5\%$). PLAN was the only other frequent function ($M=18.1\%$, $Mdn=15.1\%$). MONITOR and EVALUATE were nearly absent, with medians of $0\%$ and means of $5.7\%$ and $3.4\%$ respectively (\autoref{fig:res-dist}). This suggests that most students rarely made their comprehension needs explicit or evaluated the AI's previous responses, despite these processes being central to SRL during learning.

Among the 96 students with at least one REQUEST turn, requests were predominantly instrumental ($M=75.1\%$, $Mdn=85.7\%$) rather than executive ($M=24.9\%$). Content was primarily procedural ($M=37.4\%$) and conceptual ($M=25.1\%$), followed by answer-seeking ($M=21.1\%$) and verification ($M=16.3\%$, $Mdn=0\%$) (\autoref{fig:res-dist}). The procedural behavior matches students' stated preference for step-by-step support, but the low verification rate contrasts with the $69.7\%$ who said they wanted the AI to check their understanding.

\subsection{Intent--Enactment Alignment}
Correlations between stated intentions and corresponding enacted behaviors reveal a marked gap. The only intention significantly predicting its enacted analogue was the least socially desirable: students who selected ``just give me the final answers'' showed higher executive REQUEST rates ($46.4\%$ vs. $19.2\%$, $r=.33$, $p<.01$) and higher answer-seeking content ($39.9\%$ vs. $16.0\%$, $r=.33$, $p<.01$). Common learning-oriented intentions -- concept explanations, verification, step-by-step support -- did not significantly predict their enacted codes (all $|r|<.18$, $p>.10$). Students were thus not able to reliably translate their learning-oriented goals into aligned interaction strategies.

\subsection{Mathematics Learning Outcomes}
A paired Wilcoxon test showed a significant decrease from pre- ($67.5\%$) to post-test ($56.9\%$, $p=.014$). In an OLS model adjusting for prior math knowledge (teacher-reported), \textit{extraneous} cognitive load was the only significant predictor of post-test ($b=-.034$, $p=.025$). 

\section{Discussion}
This work-in-progress provides an initial process-sensitive account of how adolescents regulate GenAI-supported mathematics learning, by grounding their interaction characteristics into SRL and HS theory. Three patterns surface relevant for design.

First, students' stated intentions were strongly learning-oriented but their enacted behavior was dominated by instrumental requests with little explicit monitoring or evaluation. SRL theory treats monitoring and evaluation as the central mechanisms through which learners regulate understanding~\cite{Zimmerman2008InvestigatingProspects}; in our data, students often asked the AI for support but rarely made their own understanding visible, checked whether an answer addressed their goal, or asked for further clarification. Combined with the drop in learning performance, this suggests that apparently instrumental help-seeking can remain shallow when not embedded in broader SRL processes.

Second, learning-oriented intentions did not reliably translate into aligned behavior. Students frequently endorsed conceptual explanations, verification, and step-by-step support, yet these intentions only weakly predicted in-chat behavior. The only strong alignment appeared for students who explicitly chose final solutions, who were also more likely to engage in executive and answer-seeking requests. This suggests students may know which AI uses are desirable but lack the metacognitive and interactional strategies to enact them.

Third, the outcome pattern reinforces the need to examine interaction processes rather than relying on aggregate learning gains. Post-test was lower than pre-test, and higher extraneous load predicted lower post-test scores. AI tutors may thus introduce new extraneous demands related to prompt formulation, and interaction management, and can compromise meaningful learning.

The main challenge is therefore not whether students ask the AI for help, but whether they can regulate that help-seeking in ways that preserve agency, monitoring, and mathematical reasoning. This motivates our ongoing analysis of \textit{epistemic vigilance} and \textit{agency over the AI} -- two LLM-specific constructs that may capture forms of regulation classical SRL and HS codes only partly explain (whether students notice AI errors, challenge mismatched responses, impose scaffolding constraints, or actively shape the AI's tutoring role).

\section{Limitations and Future Work}
This paper reports work in progress. The behavioral analyses currently rely on AI coding alone; a stratified human-validation pass by mathematics-didactics and learning-sciences experts is ongoing, and final analyses will report per-construct reliability. The single-session design also limits conclusions about how students' AI-use strategies develop over time. Next steps include validating the codebook, coding the full dataset for epistemic vigilance and agency over the AI. 

More broadly, the results point toward GenAI learning interfaces that actively scaffold the monitoring, evaluation, and AI-shaping behaviors that adolescents may not yet bring to the interaction on their own.

\begin{credits}
\subsubsection{\ackname} We thank all the students and teachers who participated in our study study, as well as the LEAD research network for their help recruiting participants.

\subsubsection{\discintname}
This research was conducted in the absence of any commercial or financial relationship that could be construed as a potential conflict of interest.
\end{credits}

\bibliographystyle{splncs04}
\bibliography{bib}
\end{document}